\documentclass[floatfix,twocolumn,aps,prd,amsmath,amssymb]{revtex4}
\usepackage[utf8]{inputenc}
\usepackage[dvips]{graphicx}
\usepackage{dcolumn}
\usepackage{bm}
\usepackage{color}
\usepackage{ulem} 
\usepackage{url}
\usepackage{hyperref}
\usepackage{amsmath,amssymb}
\usepackage{slashed}
\usepackage{float}
\usepackage{feyn}
\usepackage{setspace}
\usepackage{wasysym}
\usepackage{xcolor}

\begin{document}

\title{Bosonic interactions in a nonlocal theory in (2+1) dimensions}

\author{Ygor Pará$^{1,2}$, Van Sérgio Alves$^{2}$, Tommaso Macrì$^{1,3}$,
E. C. Marino$^{4}$, Leandro O. Nascimento$^{5}$}

\affiliation{$^{1}$ Departamento de Física Teórica e Experimental, Universidade Federal do Rio Grande do Norte, 59072-970 Natal, Rio Grande do Norte, Brazil\\
 $^{2}$ Faculdade de Física, Universidade Federal do Pará, 66075-110 Belém, Pará, Brazil \\
 $^{3}$ International Institute of Physics, Campus Universitário
Lagoa Nova, C.P. 1613, Natal RN, 59078-970, Brazil \\
 $^{4}$ Instituto de Física, Universidade Federal do Rio de Janeiro,
C.P. 68528, Rio de Janeiro, 21941-972, Brazil \\
 $^{5}$ Faculdade de Ciências Naturais, Universidade Federal do Pará,
C.P. 68800-000, Breves, Pará, Brazil }

\date{\today}

\begin{abstract}
Pseudoquantum electrodynamics (PQED) provides an excellent description
of the interaction between charged particles confined to a plane.
When we couple a pseudogauge field with a bosonic matter field,
we obtain the so-called scalar pseudoquantum electrodynamics (SPQED).
In this work, we make a perturbative analysis of SPQED via Feynman
diagrams. We compute the one-loop Green functions: bosonic field self-energy,
electromagnetic field self-energy, and vertex corrections. Finally, we consider
the nonrelativistic interaction potential between two bosonic particles. We
compute the radiative corrections to the usual Coulomb potential and comment
on the analogies and the differences with the fermionic case.
\end{abstract}

\maketitle

\section{INTRODUCTION}

The description of electronic interactions in two-dimensional materials,
such as graphene \cite{novoselov, castroneto}, sparkled renewed interest in nonlocal relativistic quantum-field theories in reduced dimensionalities at low energies. The nonlocality in space-time is generated
by a dimensional reduction \cite{kovner, marino} that has been largely discussed in the literature \citep{VLWJF, CSBTemperature, Yukawa2018}. A well-known example is the so-called pseudoquantum electrodynamics (PQED) \cite{marino}, and sometimes, reduced quantum electrodynamics (QED) is also used in
the literature \cite{miransky}, which applies to the description of the electron-electron interactions in graphene and transition metal dichalcogenides (TMDs). In the case of graphene, quantum corrections for the longitudinal
conductivity and a driven quantum valley Hall effect are expected
to emerge at very low temperatures and a large coupling
constant \cite{PRX2015}. Furthermore, it has been shown that the correction
to the electron $g$-factor in graphene may also be calculated with this approach \cite{PRB2017}, yielding a good match with the experimental data \citep{kurganova, sing}. For TMD monolayers, conversely, the use of PQED \cite{TMDPQED} in the framework of the Bethe-Salpeter and Schwinger-Dyson formalisms has produced results for the exciton energy spectrum and lifetimes that are in excellent agreement with the experimental data \citep{hill, aslan, goerbig}. These results have been obtained making use of fermionic 
matter degrees of freedom as the particles involved in the description of 
the dynamics. However, in many cases, especially for what concerns the applications to
condensed matter systems where pairing of fermionic particles becomes relevant,
it is useful to introduce an effective bosonic description of the relevant low-energy degrees of 
freedom. 

In this paper, we study the scalar pseudoquantum electrodynamics (SPQED), which is derived from the minimal coupling between the well-known relativistic Klein-Gordon bosonic theory and a $U(1)$ gauge field in the PQED. Therefore, the model describes boson-boson interactions mediated by a nonlocal gauge field and it is renormalizable in (2+1)D. This model is related to PQED in the same way as scalar quantum electrodynamics (SQED) is connected to QED; i.e., we include a Klein-Gordon field in place of the Dirac field. Finally, we employ these results to calculate the potential among the bosonic particles, which is relevant for application to the nonrelativistic regime. It is useful to point out that the obtained nonrelativistic potential has the same form of the fermionic case, apart from a rescaling of an effective length scale, which is due to the appearance of a new diagram in the bosonic case.

The outline of this paper is as follows: In Sec.~II, we define
our model and the tree-level amplitudes. In Sec.~III, we calculate
the relevant amplitudes for renormalization at the perturbative level.
In Sec.~IV, we discuss the static potential at both
tree level and at one-loop approximation, and in Sec.~V, we draw 
the conclusions and propose some extensions of our work.
In the Appendix we review some useful identities involving the so-called Feynman parameters 
as well as standard integrals arising from the dimensional regularization.

\section{THE MODEL}

We consider the following Lagrangian 
\begin{align} 
\mathcal{L}_{\text{SPQED}} & =-\frac{1}{4}F_{\mu\nu}\left(\frac{2}{\sqrt{\Box}}\right)F^{\mu\nu}+\partial_{\mu}\phi\partial^{\mu}\phi^{\ast}-m^{2}\phi\phi^{\ast} \nonumber \\
 & +ieA_{\mu}\left(\phi\overleftrightarrow{\partial^{\mu}}\phi^{\ast}\right)+e^{2}A_{\mu}A^{\mu}\phi\phi^{\ast}+\frac{\xi}{2}A_{\mu}\frac{\partial^{\nu}\partial^{\mu}}{\sqrt{\Box}}A_{\nu},\label{Eq1}
\end{align} where $F_{\mu\nu}$ is the usual field-intensity tensor of the $U(1)$
gauge field $A_{\mu}$, which mediates the electromagnetic interaction
in 2D (pseudoelectromagnetic field), $\Box$ is the d'Alembertian
operator, $\phi$ is the massive charged Klein-Gordon field, $\phi\overleftrightarrow{\partial^{\mu}}\phi^{\ast}=\phi\left(\partial^{\mu}\phi^{\ast}\right)-\left(\partial^{\mu}\phi\right)\phi^{\ast}$, $e$ is a dimensionless coupling constant, and $\xi$ is a gauge fixing parameter.

The Feynman rules for this theory in Minkowski space are as follows.
The bare propagators are

\begin{eqnarray}
\Diagram{ & \momentum{fA}{p} & \Diagram{}
}
=S_{\phi}^{\left(0\right)}\left(p\right) & = & \dfrac{i}{p^{2}-m^{2}},\label{propScalar}\\
\Diagram{\vertexlabel_{\mu} & \momentum{gA}{p} & \Diagram{\vertexlabel_{\nu}}
}
=\Delta_{\mu\nu}^{\left(0\right)}\left(p\right) & = & \frac{-i}{2\sqrt{p^{2}}}\left[g_{\mu\nu}-\left(1-\frac{1}{\xi}\right)\frac{p_{\mu}p_{\nu}}{p^{2}}\right],\label{eq:propPQED}
\end{eqnarray}
and the vertices 
\begin{eqnarray}
\Diagram{ & fuA\vertexlabel^{p_{1}}\\
g\vertexlabel_{\mu}\\
 & fdV\vertexlabel^{p_{2}}
}
\ \ =\ \left(\Gamma_{0}^{\left(1,2\right)}\right)^{\mu}\  & = & ie\left(p_{1}+p_{2}\right)^{\mu},\label{vertice-3}\\
\Diagram{gd &  & fu\\
 & \vertexlabel_{\mu\nu}\\
gu &  & fd
}
\ \ =\ \left(\Gamma_{0}^{\left(2,2\right)}\right)^{\mu\nu} & = & ie^{2}g^{\mu\nu},
\end{eqnarray}
where we are using the notation $\Gamma^{\left(N_{A},N_{B}\right)}$
for the vertices, where $N_{A}$ and $N_{B}$ are the numbers of the gauge
field $A_{\mu}$ and bosonic scalar fields $\phi$ interacting at
the same vertex, respectively.

The nonlocal term present in the Maxwell Lagrangian in Eq. (\ref{Eq1}) renders
the canonical dimension of the gauge field equal to $1$, in units
of mass, while the scalar field has dimension $1/2$. Therefore, the
coupling constant $e$ is dimensionless in the $2+1$ space-time,
and the theory is renormalizable, analogous to scalar QED$_{4}$.

The renormalization procedure in quantum-field theory leads us to the study of the
primitively divergent Feynman diagrams.
This information can be obtained from the degree of superficial divergence
of a generic graph $\gamma$ given by 
\begin{equation}
d\left(\gamma\right)=3-N_{A}-\frac{1}{2}N_{B}.\label{eq:div-superfcial}
\end{equation}

Therefore, the quadratically divergent diagrams are those with $N_{B}=2$
and $N_{A}=0$ (scalar boson self-energy). For $N_{B}=0$ and $N_{A}=2,$
the diagrams are linearly divergent (photon self-energy), whereas
for $N_{A}=2$ and $N_{B}=2$ and for $N_{A}=1$ and $N_{B}=2,$ the
diagrams have logarithmic and linear divergence, respectively (vertex
correction). In what follows, we will calculate these diagrams using
the dimensional regularization procedure \citep{regDimone,regDimtwo,regDimthree}
as a way to obtain finite Feynman amplitudes, and we will adopt Feynman's
gauge $\xi=1$. Accordingly, the dimensionless coupling constant can
be written as $e\rightarrow e\mu^{\frac{\varepsilon}{2}},$ where
$\mu$ is an arbitrary massive parameter and $\varepsilon=3-D$.

\section{RADIATIVE CORRECTIONS}

In this section, we discuss the results of the Feynman diagrams at one loop for bosonic field self-energy (Sec. III A) as well as for the gauge field (Sec. III B).

\subsection{One-loop bosonic field self-energy}

\begin{figure}[H]
\begin{centering}
\includegraphics[width=1\columnwidth]{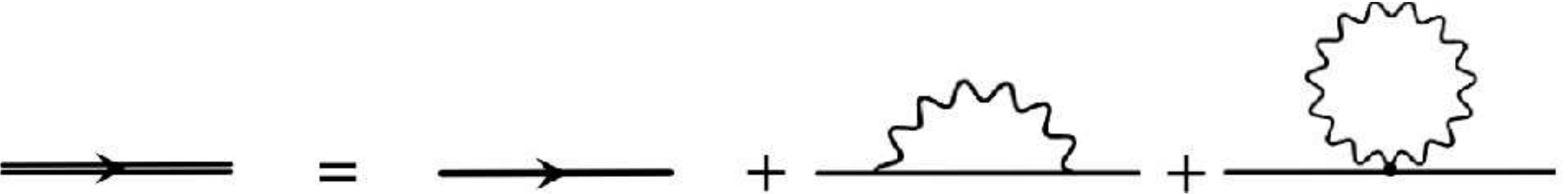} 
\par\end{centering}
\caption{The complex scalar field propagator at one-loop corrections, which
are quadratically divergent.\label{fig:camp-escalar-bosonico}}
\end{figure}

The one-loop corrections of the scalar field propagator
are shown in Fig. \ref{fig:camp-escalar-bosonico}.

\begin{figure}[H]
\begin{centering}
\includegraphics[width=0.5\columnwidth]{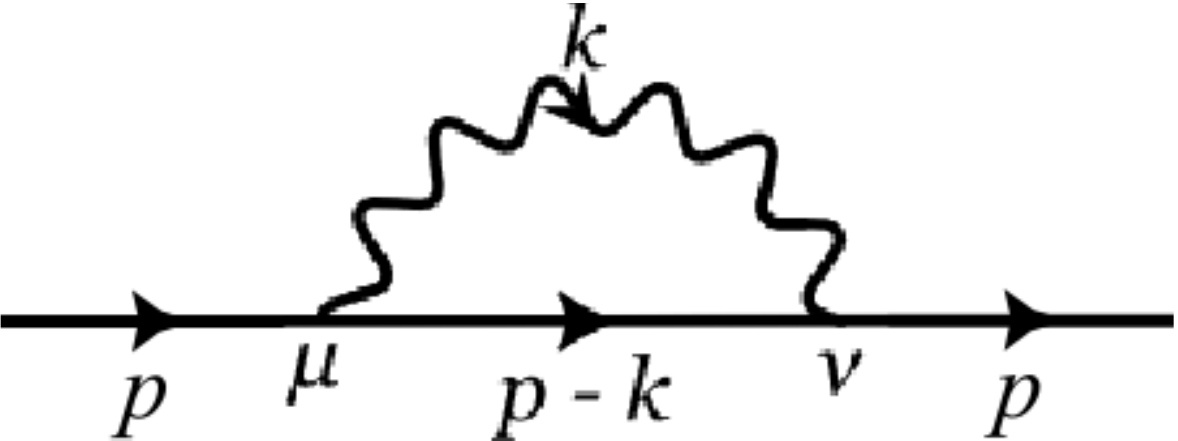} 
\par\end{centering}
\caption{One-loop correction to scalar propagator which we will call $-i\Sigma_{1}\left(p^{2},m^{2}\right).$
\label{fig:auto-energia}}
\end{figure}

We first evaluate the diagram shown in Fig. \ref{fig:auto-energia}.
We present in some detail this calculation and use the important relations
in other diagrams throughout the paper.

Using the Feynman rules (\ref{propScalar})--(\ref{vertice-3}), we have 

\begin{eqnarray}
-i\Sigma_{1}(p^{2},m^{2}) & = & \int\frac{d^{D}k}{\left(2\pi\right)^{D}}ie\mu^{\frac{\varepsilon}{2}}\left(2p-k\right)^{\mu}\Delta_{\mu\nu}^{\left(0\right)}\left(k\right)\times\nonumber \\
 &  & \times ie\mu^{\frac{\varepsilon}{2}}\left(2p-k\right)^{\nu}S_{\phi}^{\left(0\right)}\left(p-k\right).
\end{eqnarray}
We rewrite this relation as

\begin{equation}
-i\Sigma_{1}(p^{2},m^{2})=-\frac{e^{2}\mu^{\varepsilon}}{2}\int\frac{d^{D}k}{\left(2\pi\right)^{D}}\dfrac{N}{\left[\left(p-k\right)^{2}-m^{2}\right]\sqrt{k^{2}}},
\end{equation}
where $N=(2p-k)^{\mu} (2p-k)_{\mu}.$ 
To solve this integral, we use Eq.(\ref{eq:feynman-trick-ab}) in the Appendix with $\alpha=1$ and $\beta=1/2$, make a change of variable $k\rightarrow k+px,$
and use the dimensional regularization with help of Eqs. (\ref{eq:int-dim-k2}) and (\ref{eq:int-dim-1}) to obtain 
\begin{equation}
-i\Sigma_{1}(p^{2},m^{2})=\Sigma_{1}^{\text{div}}(p^{2},m^{2})+\Sigma_{1}^{\text{finite}}(p^{2},m^{2}),\label{eq:sigma1}
\end{equation}
where $\Sigma_{1}^{\text{div}}(p^{2},m^{2})$ represents the divergent
term, while $\Sigma_{1}^{\text{finite}}(p^{2},m^{2})$ is the finite
part of the diagram. Explicitly, 
\begin{eqnarray}
\Sigma_{1}^{\text{div}}(p^{2},m^{2}) & = & -\frac{e^{2}}{32\pi^{2}}\frac{1}{\varepsilon}\left(\frac{40p^{2}}{3}+8m^{2}\right),\\
\Sigma_{1}^{\text{finite}}(p^{2},m^{2}) & = & -\frac{e^{2}}{32\pi^{2}}A(p^{2},m^{2},\mu),
\end{eqnarray}
with 
\begin{align}
A(p^{2},m^{2},\mu) & =\int_{0}^{1}\frac{dx}{\sqrt{1-x}}\left[3\Delta_{1}\ln\left(\frac{4\pi\mu^{2}e^{\frac{1}{3}}}{\Delta_{1}e^{\gamma}}\right)\right.+\nonumber \\
 & +2p^{2}\left(4-4x+x^{2}\right)\ln\left.\left(\frac{4\pi\mu^{2}}{\Delta_{1}e^{\gamma}}\right)\right],\label{eq:integralA}
\end{align}
where $\Delta_{1}=p^{2}\left(x^{2}-x\right)+m^{2}x$ and $\gamma=0.5772...$
is the Euler-Mascheroni constant.

\begin{figure}[H]
\begin{centering}
\includegraphics[bb=14bp 14bp 155bp 112bp,width=0.4\columnwidth]{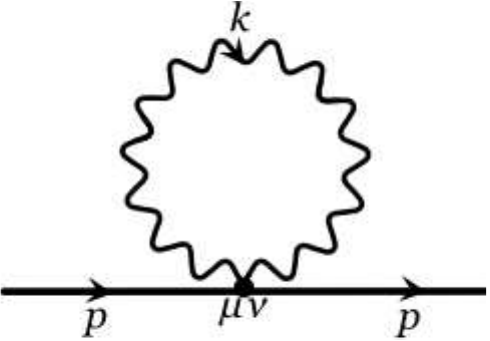}
\par\end{centering}
\caption{One-loop correction to the scalar propagator which we will call $-i\Sigma_{2}(p^{2}).$
\label{fig:autoenergia2}}
\end{figure}

The second diagram corrected the scalar field propagator and is shown
in Fig. \ref{fig:autoenergia2}. Following the Feynman rules, we
can write

\begin{equation}
-i\Sigma_{2}(p^{2})=\int\frac{d^{D}k}{\left(2\pi\right)^{D}}\left(ie^{2}\mu^{\varepsilon}g^{\mu\nu}\right)\frac{-ig_{\mu\nu}}{2\sqrt{k^{2}}}.
\end{equation}
We observe that we have an infrared divergence. To avoid this problem,
let us introduce a mass term $\widetilde{M}$, and after the
calculations, we take the limit $\widetilde{M}\rightarrow0.$ Thus,
\begin{equation}
-i\Sigma_{2}(p^{2})=\frac{3e^{2}\mu^{\varepsilon}}{2}\int\frac{d^{D}k}{\left(2\pi\right)^{D}}\frac{1}{\sqrt{k^{2}-\widetilde{M}}}.
\end{equation}
Using Eq. (\ref{eq:int-dim-1}) with Eq. (\ref{eq:log-expansion})
and taking the limits $\varepsilon\rightarrow0$ and $\widetilde{M}\rightarrow0,$
the self-energy results in \cite{Leibbrandt}
\begin{equation}
-i\Sigma_{2}(p^{2})=0.\label{eq:sigma2}
\end{equation}
Therefore, this diagram does not contribute to the perturbative series.

The one-loop corrections to the complex scalar field are given by the contributions
of Eqs. (\ref{eq:sigma1}) and (\ref{eq:sigma2}), thus, 
\begin{eqnarray}
-i\Sigma(p^{2},m^{2}) & = & -i\Sigma_{1}-i\Sigma_{2}\nonumber \\
 & = & -\frac{e^{2}}{32\pi^{2}}\frac{1}{\varepsilon}\left(\frac{40p^{2}}{3}+8m^{2}\right)\nonumber \\
 & - & \frac{e^{2}}{32\pi^{2}}A(p^{2},m^{2},\mu).\label{eq:BolhaPropEscalar}
\end{eqnarray}

After using the minimal subtraction scheme, where the divergent term
of the self-energy is neglected, we find the renormalized amplitude
given by 
\begin{equation}
-i\Sigma_{R}(p^{2},m^{2})=-\frac{e^{2}}{32\pi}A(p^{2},\mu^{2}=m^{2}),
\end{equation}
 where we have assumed $\mu^{2}=m^{2}$ for the sake of simplicity.
The pole of the full propagator, after we include the boson self-energy,
yields the renormalized mass $m_{R}$ for the bosonic field, namely,
\begin{equation}
m_{R}^{2}=m^{2}+\frac{e^{2}}{32\pi}A(p^{2}=m^{2},\mu^{2}=m^{2}).
\end{equation}
 Using Eq.(\ref{eq:integralA}), we have 
\begin{equation}
m_{R}\approx\pm|m|\sqrt{1+\frac{75\alpha}{8}}\label{eq:renormalizedmass}
\end{equation}
 with $e^{2}=4\pi\alpha$ for comparison with QED. Equation (\ref{eq:renormalizedmass})
shows that the main effect of repulsive interactions is to slightly increase
the energy gap, as expected.

\subsection{One-loop gauge field self-energy}

\begin{figure}[H]
\begin{centering}
\includegraphics[bb=14bp 14bp 167bp 92bp,width=0.55\columnwidth]{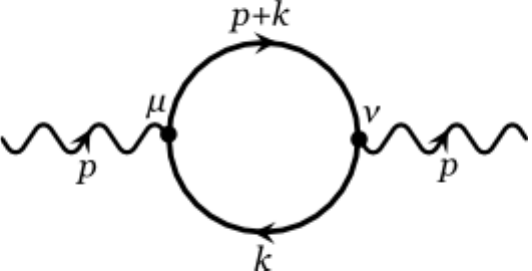}
\par\end{centering}
\caption{$d\left(G\right)=1,$ because $N_{A}=2$, $N_{B}=0$.\label{fig:Diagrama-do-tensor-polarizacao}}
\end{figure}

\begin{figure}[H]
\centering
\includegraphics[width=3cm]{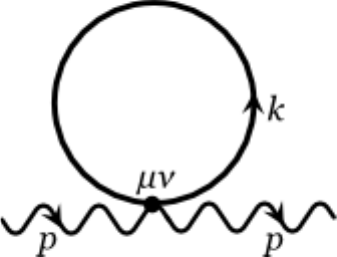}
\caption{One-loop correction to the gauge field. For this diagram, $d\left(G\right)=1$ as
$N_{A}=2$, $N_{B}=0$.
\label{fig:Diagrama-pimunu-2}}
\end{figure}

One-loop corrections to the gauge field propagator are shown in Figs. \ref{fig:Diagrama-do-tensor-polarizacao}
and \ref{fig:Diagrama-pimunu-2} and are given by 
\begin{eqnarray}
-i\Pi_{1}^{\mu\nu}(p^{2}) & = & \int\frac{d^{D}k}{\left(2\pi\right)^{D}}ie\mu^{\frac{\varepsilon}{2}}\left(2k+p\right)^{\mu}S_{\phi}^{\left(0\right)}\left(p+k\right)\times\nonumber \\
 & \times & ie\mu^{\frac{\varepsilon}{2}}\left(2k+p\right)^{\nu}S_{\phi}^{\left(0\right)}\left(k\right)
\end{eqnarray}
 and 
\begin{equation}
-i\Pi_{2}^{\mu\nu}(p^{2})=2\int\frac{d^{D}k}{\left(2\pi\right)^{D}}\left(ie^{2}\mu^{\varepsilon}g^{\mu\nu}\right)S_{\phi}^{\left(0\right)}\left(k\right).
\end{equation}

Up to that order, these diagrams are the same as $\text{SQED}_{3},$
and using dimensional regularization, we obtain 
 
\begin{align}
-i\Pi_{1}^{\mu\nu}(p^{2}) & =\frac{ie^{2}}{8\pi}\{g^{\mu\nu}\left(\sqrt{4m^{2}}-\frac{\left(p^{2}-4m^{2}\right)}{2}I\left(m,p\right)\right)\nonumber \\
 & +\frac{p^{\mu}p^{\nu}}{p^{2}}\left[\sqrt{4m^{2}}+\frac{\left(p^{2}-4m^{2}\right)}{2}I\left(m,p\right)\right]\},\label{eq:pimunu27}
\end{align}
\begin{equation}
-i\Pi_{2}^{\mu\nu}(p^{2})=-2\mu^{\varepsilon}\frac{ie^{2}}{8\pi}g^{\mu\nu}\sqrt{4m^{2}},\label{eq:pimunu2}
\end{equation}
 and therefore, 
\begin{align}
-i\Pi^{\mu\nu}(p^{2}) & =-i\Pi_{1}^{\mu\nu}(p^{2})-i\Pi_{2}^{\mu\nu}(p^{2}),\nonumber \\
 & =i\left(g^{\mu\nu}-\frac{p^{\mu}p^{\nu}}{p^{2}}\right)\pi(p^{2}),
\end{align}
 where 
\begin{equation}
\pi(p^{2})=\frac{-e^{2}}{8\pi}\left(\sqrt{4m^{2}}+\frac{(p^{2}-4m^{2})}{2}I\left(m,p\right)\right),
\label{eq:pi-pequeno-1}
\end{equation}
and 
\begin{equation}
I\left(m,p\right)=\int_{0}^{1}dx\frac{1}{\sqrt{m^{2}+p^{2}\left(x^{2}-x\right)}} ,\label{eq:i6}
\end{equation}
 which reflects the conservation of current $p_{\mu}\Pi^{\mu\nu}=0$.

Note that the use of dimensional regularization already renormalizes
the diagram \cite{Speer} which has a linear divergence in the ultraviolet regime.
The random phase approximation to the photon propagator can be visualized as 
\begin{align}
\Delta_{\mu\nu} & =\Delta_{\mu\nu}^{\left(0\right)}+\Delta_{\mu\alpha}^{\left(0\right)}\left(-i\right)\Pi^{\alpha\beta}\Delta_{\beta\nu}^{\left(0\right)}+\nonumber \\
 & +\Delta_{\mu\alpha}^{\left(0\right)}\left(-i\right)\Pi^{\alpha\beta}\Delta_{\beta\lambda}^{\left(0\right)}\left(-i\right)\Pi^{\lambda\rho}\Delta_{\rho\nu}^{\left(0\right)}+\ldots,\label{eq:serie}
\end{align}
 with $\Delta_{\mu\nu}^{\left(0\right)}=-\frac{iP_{\mu\nu}}{2\sqrt{p^{2}}},$
where $P_{\mu\nu}=g^{\mu\nu}-\frac{p^{\mu}p^{\nu}}{p^{2}}.$ We can
rewrite Eq. (\ref{eq:serie}) so that we get 
\begin{equation}
\Delta_{\mu\nu}\left(p\right)=\frac{-iP_{\mu\nu}}{2\sqrt{k^{2}}-\pi\left(p^{2}\right)}.
\end{equation} This result will be useful to obtain the static interaction potential.

\subsection{VERTEX CORRECTIONS}

\begin{figure}[H]
\begin{centering}
\includegraphics[width=1\columnwidth]{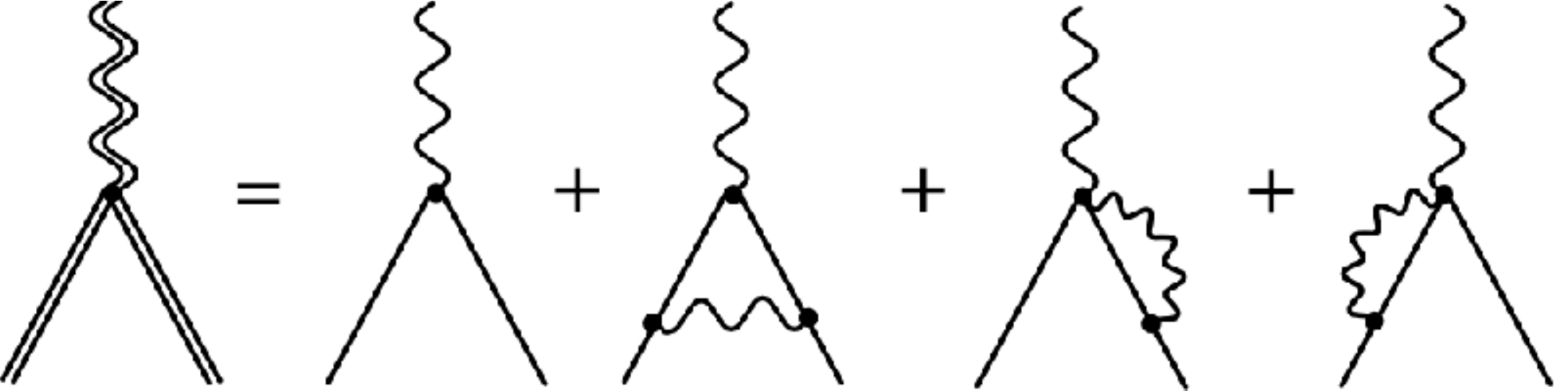}
\par\end{centering}
\caption{One-loop corrections to the vertex $ieA_{\mu}\left(\phi\protect\overleftrightarrow{\partial^{\mu}}\phi^{\ast}\right)$.
\label{fig:3-vertex}}
\end{figure}

The first vertex to be analyzed is the interaction vertex that represents
the interaction of the bosonic fields $\phi^{\ast}$ and $\phi$ with the gauge field $A_{\mu}$. The perturbative series is shown
in Fig. \ref{fig:3-vertex}.

\begin{figure}[H]
\begin{centering}
\includegraphics[width=0.3\columnwidth]{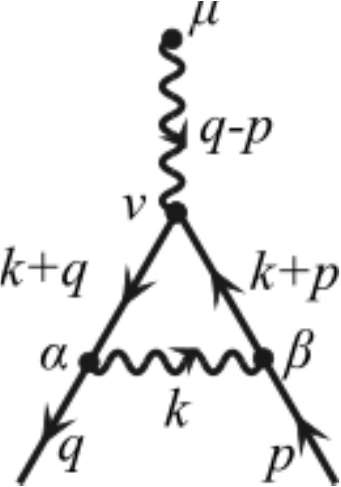} 
\par\end{centering}
\caption{One-loop correction to the vertex $ieA_{\mu}\left(\phi\protect\overleftrightarrow{\partial^{\mu}}\phi^{\ast}\right)$.
This vertex correction, $\left(\Gamma_{1}^{\left(1,2\right)}\right)^{\nu}$ is linearly divergent.\label{fig:correcao-vert-trilin}}
\end{figure}

With the use of Feynman rules (\ref{propScalar})--(\ref{vertice-3}), the diagram shown in Fig. \ref{fig:correcao-vert-trilin},
we write
\begin{align}
\left(\Gamma_{1}^{\left(1,2\right)}\right)^{\nu} & =\frac{2}{3!}\left(ie\mu^{\frac{\varepsilon}{2}}\right)^{3}\int\frac{d^{D}k}{\left(2\pi\right)^{D}}\left(2k+p+q\right)^{\nu}\times\nonumber \\
\times S_{\phi}^{\left(0\right)}\left(k+p\right) & \left(k+2p\right)^{\eta}\Delta_{\sigma\eta}^{\left(0\right)}\left(k\right)\left(k+2q\right)^{\sigma}S_{\phi}\left(k+q\right).
\end{align}
To compute the integral, we introduce the Feynman parameters through Eq. (\ref{eq:trick-abc})
with $\alpha=\beta=1$ and $\gamma=1/2$, make a change of variables
$k\rightarrow k-qx-py$, and then evaluate the integrals in $k$ variable
with Eqs. (\ref{eq:int-dim-kmuknu})--(\ref{eq:int-dim-k2}) using the expansions given by Eqs. (\ref{eq:Gamma-expansion})
and (\ref{eq:log-expansion}) and obtain 
\begin{equation}
\left(\Gamma_{1}^{\left(1,2\right)}\right)^{\nu}=\left(\Gamma_{1\left(\text{div}\right)}^{\left(1,2\right)}\right)^{\nu}+\left(\Gamma_{1\left(\text{finite}\right)}^{\left(1,2\right)}\right)^{\nu},\label{eq:VertGamma(1,2)1}
\end{equation}
with 
\begin{eqnarray}
\left(\Gamma_{1\left(\text{div}\right)}^{\left(1,2\right)}\right)^{\nu} & = & \frac{e^{3}}{12\pi^{2}}\frac{1}{\varepsilon}\left(p^{\nu}+q^{\nu}\right),\\
\left(\Gamma_{1\left(\text{finite}\right)}^{\left(1,2\right)}\right)^{\nu} & = & -\frac{e^{3}}{\pi^{2}}B^{\nu}\left(p,q,m,\mu\right),
\end{eqnarray}
with
\begin{align}
B^{\nu}\left(p,q,m,\mu\right) & =\frac{-1}{96}\int_{0}^{1}dx\int_{0}^{1-x}\frac{dy}{\sqrt{1-x-y}}\times\nonumber \\
\times\left\{ \left(U_{1}\right)^{\nu}\right. & \ln\left(\frac{4\pi\mu^{2}}{\Delta_{3}e^{\gamma}}\right)+\left(U_{2}\right)^{\nu}3\ln\left(\frac{4\pi\mu^{2}}{\Delta_{3}e^{\frac{2}{3}+\gamma}}\right)\nonumber \\
- & \left.\frac{2}{\Delta_{3}}\left(U_{3}\right)^{\nu}\right\} ,\label{eq:IntegralB}
\end{align}
where $\Delta_{3}=\left(px+qy\right)^{2}-p^{2}x-q^{2}y+m^{2}\left(x+y\right)$
and $\left(U_{1}\right)^{\nu}$, $\left(U_{2}\right)^{\nu},$ and $\left(U_{3}\right)^{\nu}$
are given by 
\begin{eqnarray}
U_{1}^{\mu} & = & q^{\mu}\left(4-4y\right)+p^{\mu}\left(4-4x\right),\label{eq:u1}\\
U_{2}^{\mu} & = & p^{\mu}\left(1-2x\right)+q^{\mu}\left(1-2y\right),\label{eq:u2}\\
U_{3}^{\mu} & = & p^{\beta}q_{\beta}p^{\mu}[\left(1-2x\right)\left(2-x\right)\left(2-y\right)+\nonumber \\
 & + & y\left(x-2x^{2}\right)]+p^{\beta}q_{\beta}q^{\mu}[\left(1-2y\right)\left(2-x\right)\times\nonumber \\
 & \times & \left(2-y\right)+yx-2y^{2}x]-p^{2}\left(2x-x^{2}\right)\times\nonumber \\
 & \times & \left[p^{\mu}\left(1-2x\right)+q^{\mu}\left(1-2y\right)\right]-q^{2}\times\nonumber \\
 & \times & [q^{\mu}\left(1-2y\right)\left(2y-y^{2}\right)+\nonumber \\
 & + & p^{\mu}\left(1-2x\right)\left(2y-y^{2}\right)].\label{eq:u3}
\end{eqnarray}

\begin{figure}[H]
\begin{centering}
\includegraphics[width=0.3\columnwidth]{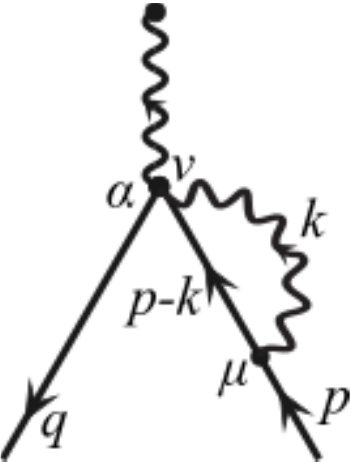} 
\par\end{centering}
\caption{One-loop correction to the vertex $ieA_{\mu}\left(\phi\protect\overleftrightarrow{\partial^{\mu}}\phi^{\ast}\right)$.
This vertex correction, $\left(\Gamma_{2}^{\left(1,2\right)}\right)^{\alpha}$
is linearly divergent.}
\label{fig:correca-vert-trilin-2}
\end{figure}

Next, we calculate the diagram in Fig. \ref{fig:correca-vert-trilin-2}.
This vertex diagram is given by 
\begin{eqnarray}
\left(\Gamma_{2}^{\left(1,2\right)}\right)^{\alpha} & = & 2\times\frac{4}{2!}\int\frac{d^{D}k}{\left(2\pi\right)^{D}}ie^{2}\mu^{\varepsilon}g^{\alpha\nu}S_{\phi}^{\left(0\right)}\left(p-k\right)\times\nonumber \\
 & \times & ie\mu^{\frac{\varepsilon}{2}}\left(2p-k\right)^{\mu}\Delta_{\mu\nu}^{\left(0\right)}\left(k\right),
\end{eqnarray}
where the first factor $2$ is due to the multiplicity of the diagram.
Introducing the Feynman parameters, Eq. (\ref{eq:feynman-trick-ab})
with $\alpha=1$ and $\beta=\frac{1}{2}$, and making the change of
variables $k\rightarrow k+xp$, we have
\begin{equation}
\Gamma_{2}^{\left(1,2\right)\alpha}=-(e\mu^{\frac{\varepsilon}{2}})^{3}p^{\alpha}\int_{0}^{1}\frac{dx\left(2-x\right)}{\sqrt{1-x}}\int\frac{d^{D}k}{\left(2\pi\right)^{D}}\frac{1}{\left[k^{2}-\Delta_{4}\right]^{\frac{3}{2}}},\label{eq:corr-vert-tril-3}
\end{equation}
with $\Delta_{4}=p^{2}\left(x^{2}-x\right)+m^{2}x.$ With the help
of Eq. (\ref{eq:int-dim-1}) and using the expansions Eq. (\ref{eq:Gamma-expansion})
and (\ref{eq:log-expansion}), we rewrite Eq. (\ref{eq:corr-vert-tril-3})
as
\begin{equation}
\left(\Gamma_{2}^{\left(1,2\right)}\right)^{\alpha}=\left(\Gamma_{2\left(\text{div}\right)}^{\left(1,2\right)}\right)^{\alpha}+\left(\Gamma_{2\left(\text{finite}\right)}^{\left(1,2\right)}\right)^{\alpha},\label{eq:VertGamma(1,2)2}
\end{equation}
where
\begin{equation}
\left(\Gamma_{2\left(\text{div}\right)}^{\left(1,2\right)}\right)^{\alpha}=-\frac{4e^{3}}{3\pi^{2}}p^{\alpha}\frac{1}{\varepsilon}
\end{equation}
and 
\begin{equation}
\left(\Gamma_{2\left(\text{finito}\right)}^{\left(1,2\right)}\right)^{\alpha}=-\frac{e^{3}}{\pi^{2}}p^{\alpha}C\left(p,m,\mu\right),
\end{equation}
with 
\begin{equation}
C\left(p,m,\mu\right)=\frac{1}{4}\int_{0}^{1}\frac{dx\left(2-x\right)}{\sqrt{1-x}}\ln\left(\frac{4\pi\mu^{2}}{\Delta_{4}e^{\gamma}}\right).\label{eq:IntegralC}
\end{equation}
Combining the results given by Eqs. (\ref{eq:VertGamma(1,2)1}) and (\ref{eq:VertGamma(1,2)2}), 
\begin{equation}
\left(\Gamma^{\left(1,2\right)}\right)^{\mu}=\left(\Gamma_{1}^{\left(1,2\right)}\right)^{\mu}+\left(\Gamma_{2}^{\left(1,2\right)}\right)^{\mu},
\end{equation}
we obtain the one-loop correction to this vertex 
\begin{eqnarray}
\left(\Gamma^{\left(1,2\right)}\right)^{\alpha} & = & -\frac{e^{3}}{\pi^{2}}\left\{ \left(\frac{5}{4}p^{\alpha}-\frac{1}{12}q^{\alpha}\right)\frac{1}{\varepsilon}+\right.\nonumber \\
 & + & \left.B^{\alpha}\left(p,q,m,\mu\right)+p^{\alpha}C\left(p,m,\mu\right)\right\} ,\label{eq:Vert(1,2)1-loop}
\end{eqnarray}
with $B^{\mu}\left(p,q,m,\mu\right)$ given by Eq. (\ref{eq:IntegralB})
and $C\left(p,m,\mu\right)$ given by Eq. (\ref{eq:IntegralC}).

\begin{figure}[H]
\begin{centering}
\includegraphics[width=1\columnwidth]{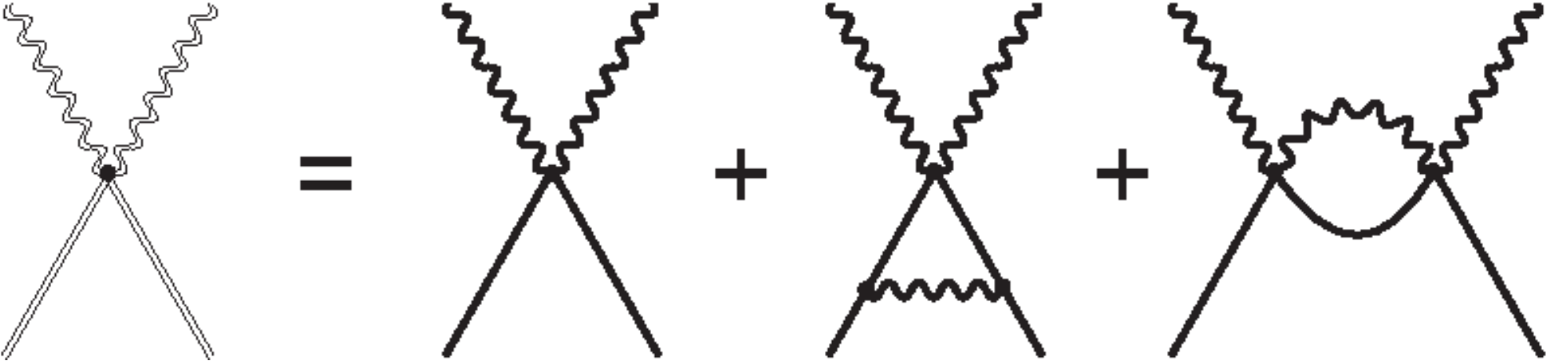} 
\par\end{centering}
\caption{One-loop corrections to the vertex $e^{2}A_{\mu}A^{\mu}\phi^{\ast}\phi$.\label{fig:corr-vert-quadrilin}}
\end{figure}

\begin{figure}[H]
\begin{centering}
\includegraphics[width=0.3\columnwidth]{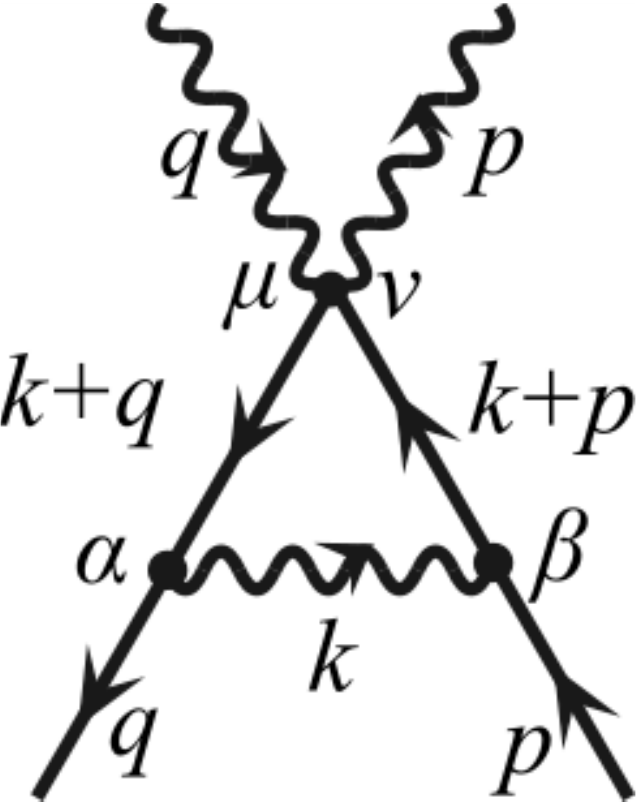} 
\par\end{centering}
\caption{One-loop correction to the vertex $e^{2}A_{\mu}A^{\mu}\phi^{\ast}\phi$.
This vertex correction $\left(\Gamma_{1}^{\left(2,2\right)}\right)^{\mu\nu}$ is logarithmically divergent.\label{fig:corr-vert-quadrilin-1}}
\end{figure}

Finally, we analyze the vertex that represents the interaction between
the scalar fields $\phi^{\ast}$ and $\phi$ with two gauge fields
$A_{\mu}$. In Fig. \ref{fig:corr-vert-quadrilin} we display the one-loop
corrections for this vertex.
First, we evaluate the diagram shown in Fig. \ref{fig:corr-vert-quadrilin-1}.

This vertex diagram is given by 
\begin{eqnarray}
\left(\Gamma_{1}^{\left(2,2\right)}\right)^{\mu\nu} & = & \frac{4}{3!}\int\frac{d^{D}k}{\left(2\pi\right)^{D}}ie^{2}\mu^{\varepsilon}g^{\mu\nu}S_{\phi}^{\left(0\right)}\left(k+q\right)\times\nonumber \\
 & \times & ie\mu^{\frac{\varepsilon}{2}}\left(k+2q\right)^{\alpha}\Delta_{\alpha\beta}^{\left(0\right)}\left(k\right)ie\mu^{\frac{\varepsilon}{2}}\left(k+2p\right)^{\beta}\times\nonumber \\
 & \times & S_{\phi}^{\left(0\right)}\left(k+p\right).
\end{eqnarray}
 Using Eq. (\ref{eq:trick-abc}), we obtain 
\begin{eqnarray}
\left(\Gamma_{1}^{\left(2,2\right)}\right)^{\mu\nu} & = & g^{\mu\nu}\frac{e^{4}\mu^{2\varepsilon}}{4}\int_{0}^{1}dx\int_{0}^{1-x}\frac{dy}{\sqrt{1-x-y}}\times\nonumber \\
\times\int & \frac{d^{D}k}{\left(2\pi\right)^{D}} & \frac{N}{\left\{ \left[k+\left(qx+py\right)\right]^{2}-\Delta_{5}\right\} ^{\frac{5}{2}}},
\end{eqnarray}
where $N=\left(k+2q\right)^{\alpha}\left(k+2p\right)_{\alpha}$ and
$\Delta_{5}=\left(qx+py\right)^{2}-q^{2}x-p^{2}y+m^{2}\left(x+y\right).$

Making the change of variables using the Eqs. (\ref{eq:int-dim-k2}) and (\ref{eq:int-dim-1}) and the expansions given by Eqs. (\ref{eq:Gamma-expansion})
and (\ref{eq:log-expansion}), we obtain 
\begin{equation}
\left(\Gamma_{1}^{\left(2,2\right)}\right)^{\mu\nu}=\left(\Gamma_{1\left(\text{div}\right)}^{\left(2,2\right)}\right)^{\mu\nu}+\left(\Gamma_{1\left(\text{finite}\right)}^{\left(2,2\right)}\right)^{\mu\nu},\label{eq:CorrVert(2,2)1}
\end{equation}
where 
\begin{eqnarray}
\left(\Gamma_{1\left(\text{div}\right)}^{\left(2,2\right)}\right)^{\mu\nu} & = & \frac{e^{4}g^{\mu\nu}}{6\pi^{2}}\frac{1}{\varepsilon},\\
\left(\Gamma_{1\left(\text{finite}\right)}^{\left(2,2\right)}\right)^{\mu\nu} & = & -\frac{e^{4}g^{\mu\nu}}{\pi^{2}}D\left(p,q,m,\mu\right),
\end{eqnarray}
and 
\begin{align}
D\left(p,q,m,\mu\right) & =-\frac{1}{48}\int_{0}^{1}dx\int_{0}^{1-x}\frac{dy}{\sqrt{1-x-y}}\times\nonumber \\
 & \times\left[3\ln\left(\frac{4\pi\mu^{2}}{\Delta_{5}e^{\gamma+\frac{2}{3}}}\right)-\frac{2f_{1}\left(p,q,x,y\right)}{\Delta_{5}}\right],\label{eq:IntegralD}
\end{align}
with 
\begin{eqnarray}
f_{1}\left(p,q,x,y\right) & = & q^{2}\left(x^{2}-2x\right)+p^{2}\left(y^{2}-2y\right)\nonumber \\
 & + & q^{\alpha}p_{\alpha}\left(2xy-2x-2y+4\right).
\end{eqnarray}

\begin{figure}[H]
\begin{centering}
\includegraphics[width=0.3\columnwidth]{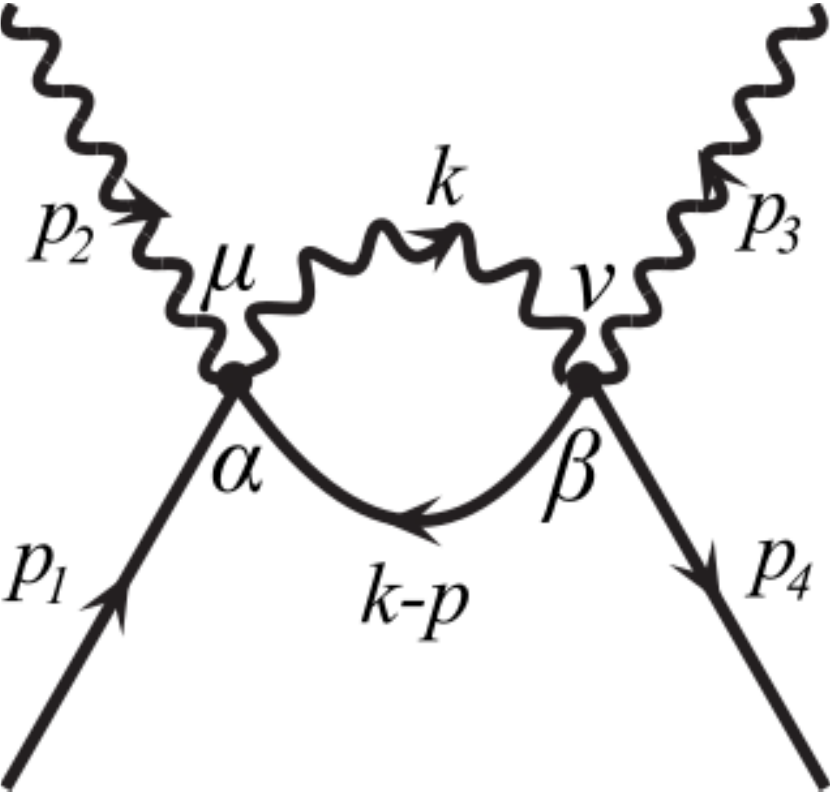}
\par\end{centering}
\caption{One-loop correction to the vertex $e^{2}A_{\mu}A^{\mu}\phi^{\ast}\phi$,
$\left(\Gamma_{2}^{\left(2,2\right)}\right)^{\mu\nu}$, which
is logarithmically divergent.\label{fig:corr-vert-quadrilin-3}}
\end{figure}

Finally, we evaluate the last diagram of the vertex correction shown
in Fig. \ref{fig:corr-vert-quadrilin-3}.

This diagram has the following analytical structure with $p=p_{1}+p_{2}$:
\begin{eqnarray}
\left(\Gamma_{2}^{\left(2,2\right)}\right)^{\mu\nu} & = & \frac{16}{2!}\int\frac{d^{D}k}{\left(2\pi\right)^{D}}\left(ie^{2}\mu^{\varepsilon}g^{\mu\alpha}\right)\Delta_{\alpha\beta}^{\left(0\right)}\left(k\right)\times\nonumber \\
 & \times & \left(ie^{2}\mu^{\varepsilon}g^{\beta\nu}\right)S_{\phi}^{\left(0\right)}\left(k-p\right).
\end{eqnarray}
Using the Feynman parametrization of Eq. (\ref{eq:feynman-trick-ab}),
with $\alpha=1$ and $\beta=\frac{1}{2}$, and further making the
change of variables $k\rightarrow k+px$, we obtain 
\begin{equation}
\left(\Gamma_{2}^{\left(2,2\right)}\right)^{\mu\nu}=-2e^{4}g^{\mu\nu}\mu^{2\varepsilon}\int_{0}^{1}\frac{dx}{\sqrt{1-x}}\int\frac{d^{D}k}{\left(2\pi\right)^{D}}\frac{1}{\left[k^{2}-\Delta_{6}\right]^{\frac{3}{2}}},
\end{equation}
with $\Delta_{6}=p^{2}\left(x^{2}-x\right)+m^{2}x$.

Solving the integral in $k$ with Eqs. (\ref{eq:int-dim-1}), (\ref{eq:Gamma-expansion}), and (\ref{eq:log-expansion}), we obtain 
\begin{equation}
\left(\Gamma_{2}^{\left(2,2\right)}\right)^{\mu\nu}=\left(\Gamma_{2\left(\text{div}\right)}^{\left(2,2\right)}\right)^{\mu\nu}+\left(\Gamma_{2\left(\text{finite}\right)}^{\left(2,2\right)}\right)^{\mu\nu},\label{eq:CorrVert(2,2)3}
\end{equation}
where 
\begin{eqnarray}
\left(\Gamma_{2\left(\text{div}\right)}^{\left(2,2\right)}\right)^{\mu\nu} & = & -\frac{2e^{4}g^{\mu\nu}}{\pi^{2}}\frac{1}{\varepsilon},\\
\left(\Gamma_{2\left(\text{finite}\right)}^{\left(2,2\right)}\right)^{\mu\nu} & = & -\frac{e^{4}g^{\mu\nu}}{\pi^{2}}E\left(p,m,\mu\right),
\end{eqnarray}
and 
\begin{equation}
E\left(p,m,\mu\right)=\frac{1}{2}\int_{0}^{1}\frac{dx}{\sqrt{1-x}}\ln\left(\frac{4\pi\mu^{2}}{\Delta_{6}e^{\gamma}}\right).\label{eq:IntegralE}
\end{equation}

Finally, combining the results of Eqs. (\ref{eq:CorrVert(2,2)1}) and (\ref{eq:CorrVert(2,2)3}) and defining 
\begin{equation}
\left(\Gamma_{\text{1-loop}}^{\left(2,2\right)}\right)^{\mu\nu}=\left(\Gamma_{1}^{\left(2,2\right)}\right)^{\mu\nu}+\left(\Gamma_{2}^{\left(2,2\right)}\right)^{\mu\nu},
\end{equation}
we obtain the result
\begin{equation}
\left(\Gamma_{\text{1-loop}}^{\left(2,2\right)}\right)^{\mu\nu}=\frac{-e^{4}g^{\mu\nu}}{\pi^{2}}\left\{ \frac{11}{6\varepsilon}+D\left(p,q,m,\mu\right)+E\left(p,m,\mu\right)\right\} .\label{eq:Vert(2,2)1-loop}
\end{equation}

\section{DISCUSSION}

To understand the effects of interactions, it is instructive to analyze the behavior of the interaction potential of SPQED in the static limit.
This potential describes how bosonic fields interact between each other. In the static limit, Eq. (\ref{eq:i6}) can be written as 
\begin{equation}
I\left(m,-\left|\mathbf{k}\right|^{2}\right)=\int_{0}^{1}\frac{dx}{\sqrt{m^{2}-\left|\mathbf{k}\right|^{2}\left(x^{2}-x\right)}}.\label{eq:int-static}
\end{equation}
 The roots of the argument are $x_{\pm}=\frac{1}{2}\pm\frac{1}{2}\sqrt{1+\frac{4m^{2}}{\left|\mathbf{k}\right|^{2}}}.$
Note that $x_{-}<0$ and $x_{+}>1$; therefore, they lie outside the integration interval.
The integral (\ref{eq:int-static}) gives 
\begin{equation}
I\left(m,-\left|\mathbf{k}\right|^{2}\right)=\frac{2}{\left|\mathbf{k}\right|}\text{arccot}\left(\frac{2m}{\left|\mathbf{k}\right|}\right).
\end{equation}
Therefore, Eq. (\ref{eq:pi-pequeno-1}) can be written as 
\begin{equation}
\pi\left(\left|\mathbf{k}\right|^{2}\right)=-\frac{\alpha}{2}\left[\sqrt{4m^{2}}-\frac{\left(\left|\mathbf{k}\right|^{2}+4m^{2}\right)}{\left|\mathbf{k}\right|}\text{arccot}\left(\frac{2m}{\left|\mathbf{k}\right|}\right)\right],\label{eq:pi-pequeno-2}
\end{equation}
 where $\alpha=e^{2}/4\pi$ is fine-structure constant. 

\begin{figure}[H]
\begin{centering}
\includegraphics[width=0.48\textwidth]{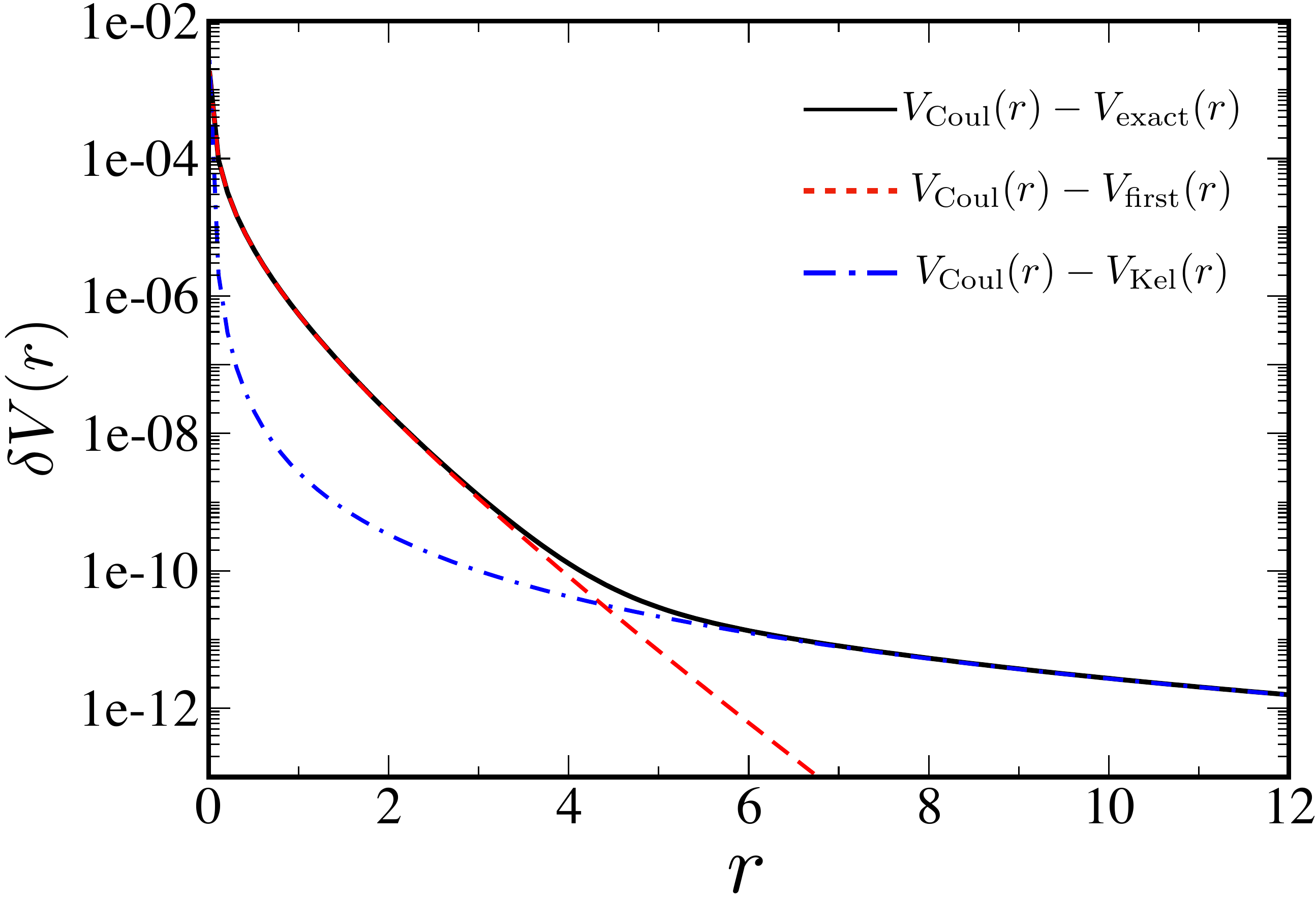}
\par\end{centering}
\caption{Differences between the potentials and Coulomb potential in different approximations of the 
one-loop result Eq.(\ref{eq:Vexact}). 
At short distances, the potential has an exponential behavior expressed in Eq. (\ref{eq:Vfirst}), typical 
of a polarization effect of the charges. At large distances, the Coulomb potential is corrected by an
isotropic dipolar potential.
\label{fig:plot}}
\end{figure}

The static interaction potential including the radiative corrections is given by
\begin{equation}
V\left(r\right)=\frac{\alpha}{\pi}\int_{0}^{\infty}\int_{0}^{2\pi}d\left|\mathbf{k}\right|d\theta\frac{\left|\mathbf{k}\right|e^{-i\left|\mathbf{k}\right|r\cos\theta}}{2\sqrt{\left|\mathbf{k}\right|^{2}}+\pi\left(\left|\mathbf{k}\right|^{2}\right)},\label{eq:potential-full}
\end{equation}
 with $r=\left|x-y\right|,$ and $\pi\left(\left|\mathbf{k}\right|^{2}\right)$
is given by Eq. (\ref{eq:pi-pequeno-2}). 
Integrating over the angular variable and defining $\left|\mathbf{k}\right|=my$,
we have 
\begin{equation}
V_{\text{exact}}\left(r\right)=\alpha m\int_{0}^{\infty}dy\frac{yJ_{0}\left(myr\right)}{y+\frac{\alpha}{4}\left\{ -2+\frac{\left(y^{2}+4\right)}{y}\text{arccot}\left(\frac{2}{y}\right)\right\} }.\label{eq:Vexact}
\end{equation}
Equation (\ref{eq:Vexact}) can be solved numerically. The result of the integration is shown by the black
continuous line in Fig. \ref{fig:plot}.

We provide an analytical estimate of the interaction potential 
at small and large distances by performing controlled approximations of 
Eq.(\ref{eq:Vexact})
We first discuss the result at small distances. Considering the first-order approximation of the Dyson 
series of the denominator
\begin{align}
V_{\text{first}}\left(r\right) & =\frac{\alpha}{r}-\frac{\alpha^{2}m}{4}\int_{0}^{\infty}dy
J_{0}\left(myr\right)\left\{ -\frac{2}{y}+\right.\nonumber \\
& +\left.\frac{\left(y^{2}+4\right)}{y^{2}}\text{arccot}\left(\frac{2}{y}\right)\right\},
\label{eq:Vfirst}
\end{align}
we expand the Bessel function at large argument (large $k$) as follows: \begin{equation}
{\displaystyle J_{0}(myr)\approx\sqrt{\frac{2}{\pi myr}}\cos\left(myr-\frac{\pi}{4}\right),}
\end{equation}
and the resulting error function as follows
\begin{equation}
\text{erf}\left(\sqrt{2}\sqrt{myr}\right)\approx\frac{2e^{-2mr}}{\sqrt{\pi}\left(\sqrt{2mr+2}+\sqrt{2mr}\right)}.
\end{equation}
Then we have
\begin{align}
V_{\text{first}}\left(r\right) & =\frac{\alpha}{r}-\frac{\alpha^{2}\sqrt{\pi}}{12}\frac{e^{-2mr}}{r}\left\{ 3\sqrt{mr+1}\right.+\nonumber \\
-\sqrt{mr} & \left.\left[8mr\left(2mr-2\sqrt{mr(mr+1)}+1\right)+1\right]\right\}.
\label{Vfirst}
\end{align}
Equation (\ref{eq:Vfirst}) is displayed in Fig. \ref{fig:plot} by the red dashed line, and shows 
that at the lowest order the interaction is well described by a Coulomb interaction with intensity
given by the fine-structure constant. The one-loop correction provides a {\it polarization}
effect coming from the virtual particle-antiparticle creation when bosonic fields approach at 
distances of the order of $r\approx 1/2m$. We point out that this effect is well known in the usual
treatment of QED in four dimensions, where similar results are found \cite{itzykson, peskin}.

To treat the large-distance behavior, we set $\left|\mathbf{k}\right|^{2}\ll4m^{2}$ in the polarization tensor.
Then, Eq. (\ref{eq:pi-pequeno-2}) can be written analytically by 
\begin{equation}
\pi_{\text{NR}}\left(\left|\mathbf{k}\right|^{2}\right)=\frac{\alpha}{6m}\left|\mathbf{k}\right|^{2}.\label{eq:pi-pequeno-non-relat}
\end{equation}
Consequently, the potential has the form of the well-known Keldysh potential \cite{K}
\begin{equation}
V_{\text{Kel}}\left(r\right)=\frac{\alpha\pi}{2r_{0}}\left[\mathbf{H}_{0}\left(\frac{r}{r_{0}}\right)-\mathrm{Y}_{0}\left(\frac{r}{r_{0}}\right)\right],\label{eq:keldysh}
\end{equation}
 where $r_{0}=\frac{\alpha}{12m}$, $\mathbf{H}_{0}\left(r/r_{0}\right),$
and $Y_{0}\left(r/r_{0}\right)$ are the Struve and the Neumann functions,
respectively. This expression has a similar structure in the
nonrelativistic regime of PQED coupled to a charged Dirac field in
$\left(2+1\right)$D \cite{LivroMarino}, with $r_{0}=2\alpha/3M$
and $M$ is the fermion mass.
To analyze the behavior of potential at long distances, we rewrite
Eq. (\ref{eq:keldysh}) in terms of a new variable $\ell=r/r_{0}$
resulting in 
\begin{equation}
r_{0}V\left(\ell\right)=\frac{\alpha\pi}{2}\left[\mathbf{H}_{0}\left(\ell\right)-\mathrm{Y}_{0}\left(\ell\right)\right].\label{eq:keldysh-1}
\end{equation}
 For $r\gg r_0,$ we can write \cite{Grads}
\begin{equation}
V\left(r\right)=\frac{\alpha}{r}-\frac{\alpha^{3}}{144m^{2}}\frac{1}{r^{3}}+\mathcal{O}\left(\frac{\alpha^{5}}{\left(12m\right)^{5}r^{5}}\right).\label{eq:keldysh-1-1}
\end{equation}
Therefore, the potential is still given by the Coulomb potential at lowest order. The first correction
is given by an isotropic dipolar potential whose strength is proportional to $\alpha^3$.

The blue line in Fig. \ref{fig:plot} shows the Keldysh potential of Eq.(\ref{eq:keldysh}), 
which captures very well the long-distance behavior of the exact result.

The above discussion of the static (nonrelativistic) potential is of direct interest in the normal phase of two-dimensional superconducting materials. 
If we add a new interaction term in Eq.(\ref{Eq1}), we have \begin{equation} \mathcal{L} = \mathcal{L}_{\text{SPQED}} - \frac{\lambda}{4!}(\phi\phi^{\ast})^2 ,  \end{equation} where the $\lambda$ parameter has a dimension of mass.  
This is a Landau-Ginzburg Lagrangian for the superconducting order parameter $\phi$ interacting with the electromagnetic potential $A_\mu$.
In the present work, we studied in detail the normal regime (above the critical temperature) where 
$m^2 \propto (T-T_c)/T_c>0$. 
In this regime, we consistently find that the interaction potential is Coulomb-like at leading order and large distances [Eq.(\ref{eq:keldysh-1-1})]. 
Below the critical temperature $m^2 < 0$ and $\lambda\neq0$, due to the Anderson-Higgs mechanism, 
the scalar field acquires a nonzero expected value in the vacuum providing a finite mass to the gauge field $A_\mu$. 
The static interaction potential $V_{A-H}(r)$ among Cooper pairs then becomes  \cite{LivroMarino} 
\begin{equation} 
V_{A-H}\left(r\right)=2 \pi \alpha \int_{0}^{\infty}
\int_{0}^{2\pi}d\left|\mathbf{k}\right|d\theta\frac{\left|\mathbf{k}\right|
e^{-i\left|\mathbf{k}\right|r\cos\theta}}{\sqrt{\left|\mathbf{k}\right|^{2}}+M},
\end{equation} 
where $M=e^{2}\left|m^{2}\right|/2\lambda$. 
Therefore,  
\begin{equation} \label{A-H}
V_{A-H}\left(r\right)=\alpha \left\{ \frac{1}{r}-\frac{\pi M}{2}
\left[\mathbf{H}_{0}\left(rM\right)-\text{Y}_{0}\left(rM\right)\right]\right\}.
\end{equation}
In the symmetry broken phase, the interaction is the sum of the Coulomb and the Keldish 
potential of Eq.(\ref{eq:keldysh}) leading effectively to a {\it screened} interaction.
Note that, at large distances, the Anderson-Higgs mechanism in SPQED has the effect of modifying the electrostatic potential to $1/r^3$. This is in contrast to the 
QED in $(3+1)$ dimensions where the interaction decays exponentially 
and has the form of a Yukawa potential \cite{LivroMarino}.
The  static potential of Eq.(\ref{A-H}) is shown 
in Fig. \ref{fig:plot} with the dot-dashed blue line.

\section{CONCLUSIONS}
In this work, we studied the scalar version of the projected quantum electrodynamics in two spatial
dimensions. We first introduced the model and defined its Feynman rules. Then we provided 
explicit calculations of the radiative corrections of the relevant interactions' vertices. Many of the
results were obtained by conveniently applying dimensional regularization to the model.
Finally, we discussed the nonrelativistic interaction between bosonic particles in the model.
We then computed the short- and large-distance behavior of such a potential. 
Whereas at short distances a polarization
effect arises, similar to the four-dimensional QED, at large distances the Coulomb potential gets 
modified by an isotropic dipolar potential.
These results are of importance for a series of applications in condensed matter physics, where a 
bosonic description might be more adequate than the correspondent fermionic counterpart. We discussed the applicability of these results to the normal and superconducting 
phase of thin superconducting films \cite{marino-hans}. Extensions of this work with the inclusion of self-interacting bosonic fields ($\lambda\neq 0$) and to other platforms such as exciton-polariton condensation, or the quantum simulation
of analog models of relativistic field theories with ultracold atoms will be considered elsewhere.

\section{ACKNOWLEDGEMENTS}
This study was financed in part by the Coordena\c{c}\~{a}o de Aperfei\c{c}oamento de Pessoal de N\'{i}vel Superior, Brasil, Finance Code 001. V. S. A. is supported in part by CNPq. T.M. acknowledges
CNPq for support through Bolsa de produtividade em Pesquisa Grant No. 311079/2015-6. E.C.M. is supported in part by CNPq and FAPERJ.
This work was supported by the Serrapilheira Institute (Grant No. Serra-1812-27802 to T.M.), 
CAPES-NUFFIC Project No. 88887.156521/2017-00. The authors also thank G. C. Magalh\~{a}es, R. F. Ozela, and L. F. F. Aguilar for useful discussions.

\appendix

\section{USEFUL IDENTITIES}
In this appendix we review some useful identities involving the so-called Feynman parameters 
often encountered in QED calculations \cite{peskin} 
that also appear in our one-loop calculations discussed
in the main text as well as integrals arising from dimensional regularization.

We start with the following two identities:
\begin{equation}
\frac{1}{a^{\alpha}b^{\beta}}=\frac{\Gamma\left(\alpha+\beta\right)}{\Gamma\left(\alpha\right)\Gamma\left(\beta\right)}\int_{0}^{1}dx\frac{x^{\alpha-1}\left(1-x\right)^{\beta-1}}{\left[ax+b\left(1-x\right)\right]^{\alpha+\beta}},\label{eq:feynman-trick-ab}
\end{equation}
\begin{align}
\frac{1}{a^{\alpha}b^{\beta}c^{\gamma}} & =\frac{\Gamma\left(\alpha+\beta+\gamma\right)}{\Gamma\left(\alpha\right)\Gamma\left(\beta\right)\Gamma\left(\gamma\right)}\times\nonumber \\
\times\int_{0}^{1} & dx\int_{0}^{1-x}dy\frac{\left(1-x\right)^{\alpha-1}y^{\gamma-1}\left(x-y\right)^{\beta-1}}{\left[ax+by+c\left(1-x-y\right)\right]^{\alpha+\beta+\gamma}}.\label{eq:trick-abc}
\end{align}

When performing dimensional regularization, we make use of the integrals
\begin{equation}
\int\frac{d^{D}k}{\left(2\pi\right)^{D}}\frac{k^{2}}{\left[k^{2}-\Delta\right]^{\alpha}}=\frac{i^{2\alpha-1}}{\left(4\pi\right)^{\frac{D}{2}}}\frac{D}{2}\frac{\Gamma\left(\alpha-\frac{D}{2}-1\right)}{\Gamma\left(\alpha\right)}\Delta^{\frac{D}{2}+1-\alpha},\label{eq:int-dim-k2}
\end{equation}
\begin{equation}
\int\frac{d^{D}k}{\left(2\pi\right)^{D}}\frac{1}{\left[k^{2}-\Delta\right]^{\alpha}}=\frac{i^{2\alpha+1}}{\left(4\pi\right)^{\frac{D}{2}}}\frac{\Gamma\left(\alpha-\frac{D}{2}\right)}{\Gamma\left(\alpha\right)}\Delta^{\frac{D}{2}-\alpha},\label{eq:int-dim-1}
\end{equation}
\begin{equation}
\int\frac{d^{D}k}{\left(2\pi\right)^{D}}\frac{k^{\mu}k^{\nu}}{\left[k^{2}-\Delta\right]^{\alpha}}=\frac{g^{\mu\nu}}{D}\int\frac{d^{D}k}{\left(2\pi\right)^{D}}\frac{k^{2}}{\left[k^{2}-\Delta\right]^{\alpha}},\label{eq:int-dim-kmuknu}
\end{equation}
where $\Delta$ is a parameter with the dimension of a mass.
Finally, it is useful to recall the expansion of the gamma function for a small argument
\begin{equation}
\Gamma\left(\frac{\varepsilon}{2}\right)=\frac{2}{\varepsilon}-\gamma+\mathcal{O}\left(\varepsilon\right),\label{eq:Gamma-expansion}
\end{equation}
where $\gamma=0.5772...$ is the Euler-Mascheroni constant. Also, for an arbitrary quantity $Q$ in the limit
of vanishing $\varepsilon$, we have
\begin{equation}
Q^{\frac{\varepsilon}{2}}\simeq1+\frac{\varepsilon}{2}\ln Q.\label{eq:log-expansion}
\end{equation}

\end{document}